\documentclass[preprint,aps,draft]{revtex4}

\usepackage{graphicx}
\usepackage{dcolumn}
\usepackage{bm}

\begin{document}

\title{Dynamical Theory of Disoriented Chiral Condensates 
at QCD Phase Transition\footnote{Talk at Compact Stars: Quest For New 
States of Dense Matter, KIAS-APCTP International Symposium 
in Astro-Hadron Physics, 2003.}}

\author{Sang Pyo Kim}\email{sangkim@kunsan.ac.kr}

\affiliation{Department of Physics, Kunsan National University,
Kunsan 573-701, Korea}

\date{\today}
\begin{abstract}
We apply the canonical quantum field theory based on the 
Liouville-von Neumann equation to the nonequilibrium linear sigma model.
Particular emphasis is put on the mechanism for domain growth of 
disoriented chiral condensates
due to long wavelength modes and its scaling behavior.
Scattering effects, decoherence and emergence of order parameter are 
also discussed beyond the Hartree approximation.
\end{abstract}

\maketitle

\section{Introduction}

The high density and temperature state of hadronic matter consists of  
quark-gluon plasma and would have occurred in the
early universe or may be realizable in heavy ion collision 
experiments. In QCD with two massless quarks, the chiral symmetry
$SU(2)_L \times SU(2)_R$ at high temperatures spontaneously breaks down 
to $SU(2)_{L+R}$ at lower temperatures by 
the quark-antiquark condensate
$\langle {\bar q}_L^i q_{Rj} \rangle = \sigma \delta^i_j + i \vec{\pi} \cdot 
\vec{\tau}^i_j$, an order parameter. 
The field $\phi_a = (\sigma, \vec{\pi})$ respects $O(4)$ rotations and thus belongs 
to the universality class of a four component isotropic 
Heisenberg antiferromagnet \cite{wilczek}. The effective theory of 
the QCD phase transition is described by the linear sigma model \cite{rajagopal} 
or equally by a mean-field theory of Polyakov loops  
and/or by the glueball fields for the hadronic states of QCD \cite{pisarski}.

In QCD the $SU(2)$ phase transition would probably be a second order 
whereas the $SU(3)$ phase transition would be a weakly first order. 
In relativistic heavy ion collisions, hot and high dense regions 
are made, where the chiral symmetry would be restored, and as the regions cool, 
the quark-gluon plasma would undergo the second order phase transition. In second 
order phase transitions, as temperature approaches a critical temperature, 
the correlation length cannot grow indefinitely 
and must be frozen due to causality \cite{kibble}. 
However, the cooling process may be fast enough for the kinematic
time scale of quench to be smaller than thermal relaxation time scale. 
The rapid expansion of the quark-gluon plasma enforces 
a rapid quench and results in a phase transition 
far from equilibrium (nonequilibrium). 
In such a nonequilibrium phase transition domains 
(regions of misaligned vacuum) develop due to the instability of long 
wavelength modes.

Another possible candidate for the QCD phase transition is Centauri events
in comic rays \cite{lattes}.
Anomalously large event-by-event fluctuations have been observed
in the ratio of charged to neutral pions
and thus require a new theory or interpretation.
The disoriented chiral condensate (DCC) of classical pion fields was introduced as 
one of the proposed mechanisms for coherent emission of pions from a large
domain \cite{anselm}. If the QCD phase transition proceeds in equilibrium, 
all directions of $\vec{\pi}$ are equally probable and domains 
of size $1/T_c$  cannot explain 
the anomalous production of pions of a certain kind. 
Rajagopal and Wilczek advocated the nonequilibrium QCD
phase transition through a quench for DCC domain growth, where 
unstable long wavelength modes of pions are exponentially amplified 
\cite{rajagopal}.

In this talk, we adopt a recently introduced canonical field method 
to elaborate the nonequilibrium QCD phase transition based on 
the linear sigma model. The quench process is imitated by a mass squared 
that changes signs during a finite quench time \cite{kim1,kim2,kim-khanna}. 
We particularly focus on the 
dynamical process of domain formation from long wavelength modes of pions growing 
exponentially during an instability period and on the scaling behavior of 
domains size. The scaling behavior of domains has been found through simulations 
in condensed matter systems and cosmology. We further discuss the effects of 
direct scatterings among each pion field modes 
on domains size and on decohering long wavelength 
modes and emergence of an order parameter.

\section{Nonequilibrium Linear Sigma Model}

The effective theory of QCD phase transition with two massless quarks is 
described by the quark-antiquark condensate $\langle \bar{q} q \rangle$, 
which, in turn, defines  $\phi_a = (\sigma, \vec{\pi})$, $\vec{\pi}$ 
being the pion field. The quark mass $m_q$ provides a symmetry breaking 
external field to an, otherwise, $O(4)$ symmetric field theory 
at high temperatures.
To include a quench, we may write the linear sigma model \cite{gellmann}
in the form
\begin{equation}
L = \int d^3 x \biggl[ \frac{1}{2} \partial_{\mu} 
\phi^a \partial^{\nu} \phi_a - \frac{\lambda}{4}( 
\phi^a \phi_a)^2 - \frac{1}{2} m^2(t) \phi^a \phi_a 
+ H \sigma \biggr]. \label{noneq sigma}
\end{equation}
The time-dependent quench process has been imitated by the 
symmetry breaking mass squared $m^2(t)$, which changes signs 
during the phase transition. In this sense Eq. (\ref{noneq sigma}) is 
the nonequilibrium linear sigma model.
As $H$ does not affect much  the instability of long wavelength modes 
during the phase transition, we assume $ H = 0$ for simplicity reason.

To illustrate how nonequilibrium phase transitions in general 
affect formation of domains, we consider a simple field model 
with a quench time scale \cite{kim1}.
The simple model motivated by the nonequilibrium linear sigma model
is given by the potential
\begin{equation}
V (\phi_a) = \frac{m^2 (t)}{2} \phi^a \phi_a + \frac{\lambda}{4} 
(\phi^a \phi_a)^2,
\label{phi4}
\end{equation}
with the mass squared
\begin{equation}
m^2 (t) = m_1^2 - m_0^2 \tanh \Bigl(\frac{t}{\tau}\Bigr),
\quad (m_0^2 > m_1^2) \label{quen}.
\end{equation}
In the limit of zero quench time $(\tau = 0)$, we obtain the instantaneous
(sudden) quench. Now finite temperature field theory \cite{dolan} 
cannot be applied to this model when the quench time scale
$\tau$ is smaller than relaxation time scale.
Further, in the second order phase transitions, 
long wavelength modes grow exponentially during rolling over the barrier. 
Finite temperature field theory does not properly 
take into account dynamical processes of phase transitions.

One therefore needs some nonequilibrium quantum field theory 
when the kinematic time scale is smaller than  
thermal relaxation time scale so that finite temperature field theory 
cannot be applied.
There are several methods for nonequilibrium quantum fields such as 
the closed-time path integral \cite{schwinger} and 
the functional Schr\"{o}dinger-picture \cite{freese}. 
Recently there has 
been developed a canonical method based on the quantum 
Liouville-von Neumann equation, which provides all time-dependent Fock states 
at the leading order \cite{kim1,kim2}. The new 
canonical method is equivalent, at leading order, to the 
time-dependent Hartree approximation,
but it can go beyond the Hartree approximation since any perturbation method
can be readily applied to these Fock states.

As a nonlinear theory, the linear sigma model has defied yet any 
nonperturbative solution in a closed form. We can, at best, rely on 
perturbation methods. The Hartree approximation, though being 
a perturbation scheme, includes some part of nonperturbative effects 
at the lowest order \cite{boyanovsky,cooper}.    
In the Hartree approximation, dividing the $\sigma$ field into a background 
field $\phi(t)$ and its quantum fluctuation $\chi ({\bf x}, t)$, 
and using the Hartree factorization \cite{boyanovsky}, we obtain
the truncated Hamiltonian for the linear sigma model
\begin{equation}
H_0  = \int d^3x \biggl[ \frac{\pi_{\chi}^2}{2} 
+ \frac{\pi_{\vec{\pi}}^2}{2} + \frac{(\nabla \chi)^2}{2} 
+ \frac{(\nabla \vec{\pi})^2}{2} + h_{\phi}(t) \chi + \frac{m^2_{\chi} (t)}{2}
\chi^2 + \frac{m_{\vec{\pi}}^2}{2} \vec{\pi}^2 \biggr], \label{har ham}
\end{equation}
where the effective couplings are
\begin{eqnarray}
h_{\phi}(t) &=& \phi (t) [m^2 (t) + 4 \lambda \phi^2 (t) + 4 \lambda \langle
\vec{\pi}^2  \rangle ], \nonumber\\
m_{\chi}^2 (t) &=& m^2 (t) + 4 \lambda \phi^2 (t) + 4 \lambda \langle
\vec{\pi}^2 \rangle, \nonumber\\
m_{\vec{\pi}}^2 (t) &=& m^2 (t) + 12 \lambda \phi^2 (t) + 4 \lambda \langle
\vec{\pi}^2 \rangle.
\end{eqnarray}

\section{DCC Domain Growth}

Quantum dynamics of DCC can be further approximated 
by an exactly solvable model now motivated 
by the linear sigma model (\ref{noneq sigma}) or 
the Hartree approximation (\ref{har ham}).
The model Hamiltonian for the pion field $\phi_a$ is \cite{amado}
\begin{equation}
H_0(t) = \int d^3 x \Bigl[\frac{1}{2} \pi^a \pi_a + \frac{1}{2} (\nabla 
\phi^a \cdot \nabla \phi_a)^2
+ \frac{1}{2} m^2 (t) \phi^a \phi_a \Bigr]. \label{mod ham}
\end{equation}
Note that all pion fields are decoupled from each other since 
the nonlinear term is neglected at this moment. 
In terms of the Fourier cosine and sine modes
\begin{equation}
\phi_{\bf k}^{(+)} (t) = \frac{1}{2} [\phi_{\bf k} (t) + \phi_{-
{\bf k}} (t)], \quad \phi_{\bf k}^{(-)} (t) =
\frac{i}{2}[\phi_{\bf k} (t) - \phi_{- {\bf k}} (t)],
\end{equation}
and with a compact notations $\alpha = \{ (\pm), {\bf k} \}$,
the Hamiltonian becomes a sum of decoupled time-dependent oscillators
\begin{equation}
H_0(t) = \sum_{a \alpha} \frac{1}{2} \pi_{a \alpha}^2 + \frac{1}{2} 
\omega_{\alpha}^2 (t) \phi_{a \alpha}^2, \quad (\omega_{\alpha}^2 (t) = {\bf k}^2 
+ m^2 (t)). \label{osc ham}
\end{equation}
Then quantum states of the pion field are found by the time-dependent
creation and annihilation operators \cite{kim1,kim2}
\begin{eqnarray}
\hat{a}_{a \alpha}^{\dagger} (t)  &=& -i [ \varphi_{a \alpha} (t)
\hat{\pi}_{a \alpha} - \dot{\varphi}_{a \alpha} (t)
\hat{\phi}_{a \alpha}], \nonumber\\ \hat{a}_{a \alpha} (t) & =& i [
\varphi_{a \alpha}^* (t) \hat{\pi}_{a \alpha} -
\dot{\varphi}^*_{a \alpha} (t) \hat{\phi}_{a \alpha}], \label{cr an}
\end{eqnarray}
where $\hat{\pi}_{a \alpha}$ and $\hat{\phi}_{a \alpha}$ are 
Schr\"{o}dinger operators.
Note that these operators do not diagonalize the Hamiltonian (\ref{osc ham}), 
but satisfy the Liouville-von Neumann equations
\begin{eqnarray}
i \frac{\partial}{\partial t} \hat{a}^{\dagger}_{a \alpha} (t) +
[\hat{a}^{\dagger}_{a \alpha} (t), \hat{H}_{a \alpha} (t)] = 0, \quad
i \frac{\partial}{\partial t} \hat{a}_{a \alpha} (t) +
[\hat{a}_{a \alpha} (t), \hat{H}_{a \alpha} (t)] = 0,
\end{eqnarray}
which lead to mean-field equations for the auxiliary field variables
\begin{equation}
\ddot{\varphi}_{a \alpha} (t) + \omega_{\alpha}^2 (t)
\varphi_{a \alpha} (t) = 0. \label{aux eq}
\end{equation}
In fact, these operators satisfy the equal time commutation relations
\begin{equation}
[ \hat{a}_{a \alpha} (t), \hat{a}^{\dagger}_{b \beta} (t) ] =
\delta_{ab} \delta_{\alpha \beta},
\end{equation}
with the aid of the Wronskian condition
\begin{equation}
\dot{\varphi}^*_{a \alpha} \varphi_{a \alpha} -
\varphi^*_{a \alpha} \dot{\varphi}_{a \alpha} = i. \label{wr con}
\end{equation}

The Fock space for each pion field mode consists of number states 
defined as
\begin{equation}
\hat{N}_{a \alpha} (t) \vert n_{a \alpha}, t \rangle_0 \equiv
\hat{a}^{\dagger}_{a \alpha} (t) \hat{a}_{a \alpha} (t) \vert
n_{a \alpha}, t \rangle_0 = n_{a \alpha} \vert n_{a \alpha}, t
\rangle_0.
\end{equation}
We should note that these are exact quantum states 
of the time-dependent Schr\"{o}dinger
equation for the Hamiltonian (\ref{mod ham}) or (\ref{osc ham}). 
The quantum state of the pion field itself is then a product
of each mode state for each pion field. 
Of a particular interest is the Gaussian vacuum state of the pion
field
\begin{equation}
\vert 0, t \rangle_0 = \prod_{a \alpha} \vert 0_{a \alpha}, t
\rangle_0. \label{gauss vac}
\end{equation}
Now the Green function for the pion field is simply given by
\begin{equation}
G_0 ({\bf x}, t; {\bf x}', t' \rangle = \prod_{a \alpha} G_{0
a \alpha} (\phi_{a \alpha}, t; \phi_{a \alpha}', t'), \label{gr fn1}
\end{equation}
where the $(a \alpha)$-mode Green function takes the form
\begin{equation}
G_{0 a \alpha} (\phi_{a \alpha}, t; \phi_{a \alpha}', t')  =
\sum_{n_{a \alpha}} \langle \phi_{a \alpha} \vert n_{a \alpha}, t
\rangle_0 ~{}_0\langle n_{a \alpha}, t' \vert \phi_{a \alpha}'
\rangle.
\end{equation}

To study formation of domains during a nonequilibrium quench process, 
we consider the smooth finite quench (\ref{quen}). 
The free field theory (\ref{mod ham}) 
is then exactly solvable \cite{kim1}, and is a good 
approximation for the linear sigma model as long as
$|m^2(t)|$ is larger than $|\phi(t)|$ and $\langle
\vec{\pi} \rangle^2 = \sum_{b \beta} \varphi_{b \beta}^*
\varphi_{b \beta}$.   
Far before the phase transition, each mode is stable and oscillates 
around the true vacuum with the solution
\begin{equation}
\varphi_{a \alpha i} (t) = \frac{1}{\sqrt{2 \omega_{\alpha i}}}
e^{- i \omega_{\alpha i} t}, \quad \omega_{\alpha i} = \sqrt{{\bf
k}^2 + m^2_i}.
\end{equation}
The two-point correlation function for each component of pion field
is the Green function at equal
times
\begin{equation}
G_{0 a} ({\bf x}, {\bf x}', t) = \langle
\hat{\phi}_a ({\bf x}, t)
\hat{\phi}_a ({\bf x}', t) \rangle_0 = G_{0 a} ({\bf x}, t; {\bf x}', t)
\end{equation}
with respect to the Gaussian vacuum or thermal equilibrium.

On the other hand, after the phase transition  $(m^2 = - m_f^2 = 
- (m_0^2 - m_1^2))$,
the long wavelength modes with $k < m_f$ become unstable
and exponentially grow, whereas short wavelength modes with
$k > m_f$ are still stable and oscillate around the false vacuum.
Far after the phase transition, the unstable long wavelength
modes have the asymptotic solutions
\begin{equation}
\varphi_{a \alpha f} = \frac{\mu_{\bf k}}{\sqrt{2} (m_f^2 - {\bf
k}^2)^{1/2}} e^{(m_f^2 - {\bf k}^2)^{1/2} t} + \frac{\nu_{\bf
k}}{\sqrt{2} (m_f^2 - {\bf k}^2)^{1/2}} e^{- (m_f^2 - {\bf
k}^2)^{1/2} t}.
\end{equation}
Here $\mu_{\bf k}$ and $\nu_{\bf k}$ depend on the
quench process and satisfy the relation $|\mu_{\bf k}|^2 -
|\nu_{\bf k}|^2 = 1$. The correlation function is then
dominated by the exponentially growing part
\begin{equation}
G_{0 a f} ({\bf x}, t; {\bf x}', t) \simeq \int_0^{m_f} \frac{d^3
{\bf k}}{(2 \pi)^3} e^{i {\bf k} \cdot ({\bf x} - {\bf x}' )}
|\mu_{\bf k}|^2 \frac{e^{2 \sqrt{m_f^2 - {\bf k}^2} t}}{2 ({\bf
k}^2 - m_f^2)}. \label{2 fn}
\end{equation}

Using the exact solutions \cite{kim1}, 
we obtain the two-point thermal correlation function for each pion field 
at the intermediate stage of the quench $( - \tau < t < \tau)$
\begin{equation}
G_{0 a T} (r,t) \simeq G_{0 a T} (0, t) \frac{\sin \Bigl(
\frac{\sqrt{\tau t}}{m_0} r \Bigr)}{\frac{\sqrt{\tau t}}{m_0} r}
\exp \Bigl( - \frac{r^2}{8 \frac{\sqrt{\tau t}}{m_0}} \Bigr).
\end{equation}
It follows that domains obey a scaling relation
for the correlation length
\begin{equation}
\xi_D (t) = 2 \Bigl(\frac{2 \tau t}{m_0^2} \Bigr)^{1/4}. \label{scal rel1}
\end{equation}
The power $1/4$ has been found in numerical simulations \cite{holzwarth}.
At the later stage far after the quench $(t \gg \tau)$, 
domains still show the Cahn-Allen scaling behavior but with 
a different power
\begin{equation}
\xi_D (t) = 2 \Bigl(\frac{2 \tilde{t}}{m_f} \Bigr)^{1/2}, 
\quad \tilde{t} = t - \frac{\tau^3}{8}
[\zeta(3) - 1] (m_i^2 + m_f^2). \label{scal rel2}
\end{equation}
The scaling relation for the instantaneous
quench is obtained by letting $\tau = 0$ in Eq. (\ref{scal rel2}).
A kind of resonance has also been observed in the correlation function with
simple poles at
\begin{equation}
\tau = \frac{n}{(m_f^2 - {\bf k}^2)^{1/2}}, ~~ (n = 1, 2, 3, \cdots).
\end{equation}
This structure implies certain adjusted quench rates $\tau$ may lead to sufficiently 
large domains \cite{kim1}.

\section{Scattering, Decoherence and Order Parameter}

The tree level approximation in Sec. 3 does take into account neither any 
interaction among different modes of each pion field nor 
the interaction between pions. Similarly the 
Hartree approximation includes only mean-field effects among modes and 
pion fields and  thus neglects any direct scatterings among modes. 
To go beyond the Hartree 
approximation we may use the formalism in Ref. 8. The 
wave functional for the Schr\"{o}dinger equation
can be expressed in terms of the Green function (kernel or
propagator) as
\begin{equation}
\Psi ({\bf x}, t) = \int G({\bf x}, t; {\bf x}_0, t_0) \Psi({\bf
x}_0, t_0) d{\bf x}_0 dt_0.
\end{equation}
As the linear sigma model is nonlinear, we use a perturbation method. 
We divide the Hamiltonian 
\begin{equation}
H (t) = H_0 (t) + \lambda H_P (t),
\end{equation} 
into an exactly solvable (quadratic) Gaussian and a perturbation part
\begin{eqnarray}
H_0 &=& \frac{1}{2} \pi_{\phi}^a \pi_{\phi a} 
+ \frac{1}{2} (\nabla \phi^a \cdot \nabla \phi_a)^2 + \frac{1}{2} 
( m^2 + 9 \lambda \langle \phi^b \phi_b \rangle ) \phi^a \phi_a, \nonumber\\
H_P &=& \frac{1}{4} (\phi^a \phi_a)^2 - \frac{9}{2} \langle \phi^b
\phi_b \rangle \phi^a \phi_a.
\end{eqnarray}
Then in terms of the Green function for $\hat{H}_0$ in Sec. 3,
\begin{equation}
\Bigl(i \frac{\partial}{\partial t} - \hat{H}_0 ({\bf x}, t)
\Bigr) G_0 ({\bf x}, t; {\bf x}', t') = \delta ({\bf x} - {\bf
x}') \delta (t - t'), \label{gr fun}
\end{equation}
we write the wave functional as
\begin{equation}
\Psi ({\bf x}, t) = \Psi_0 ({\bf x}, t) + \lambda \int G_0 ({\bf
x}, t; {\bf x}', t') \hat{H}_P ({\bf x}', t') \Psi ({\bf x}', t')
d{\bf x}' dt', \label{gr wav1}
\end{equation}
and finally obtain the wave functional of the form
\begin{eqnarray}
\Psi (1) &=& \Psi_0 (1) + \lambda \int G_0 (1,2) \hat{H}_P (2)
\Psi_0 (2) \nonumber\\ && 
+ \lambda^2 \int \int G_0 (1,2) \hat{H}_P (2) G_0 (2,3)
\hat{H}_P (3) \Psi_0 (3) + \cdots, \label{gr wav2}
\end{eqnarray}
where $(i)$ denotes $({\bf x}_i, t_i)$ and 
$\Psi_0$ is a wave functional for $\hat{H}_0$.
This method goes beyond the Hartree approximation and
gives us the wave functional in a series of $\lambda$
\begin{equation}
\Psi ({\bf x}, t) = \Psi_0 ( {\bf x}, t ) + \sum_{n = 1}
\lambda^n \Psi_0^{(n)}. \label{wav sum}
\end{equation}
The term $\Psi_0^{(n)}$ comes from $n$th order correction of $H_P$.

We now give a few remarks on the effects on DCC formation 
of the non-Gaussian (beyond the Hartree) approximation. 
First, the nonlinear correction due to $\Psi^{(n)}_0$  
enhances the correlation length for domains by a factor \cite{kim-khanna}
\begin{equation}
\frac{\xi_{Dn}}{\xi_D} = (2 n +1)^{1/2}.
\end{equation}
Higher order correction terms begin to grow 
provided that the duration of instability is long enough before crossing the 
inflection point. This condition may be provided by the quench time scale
that is comparatively large but still smaller 
than relaxation time scale. Under 
this condition, significantly large DCC domains
may lead to observable effects in heavy ion collisions 
or high energy cosmic rays. 
Second, the higher order correction terms in 
Eq. (\ref{wav sum}) include direct scattering effects. The direct 
scattering can be shown obviously in Eq. (\ref{gr wav2}), where the Green function 
is simply given by
\begin{equation}
G_0 ({\bf x}, t; {\bf x}', t') = \sum_{q = 0}^{\infty} \Psi_q ({\bf x}, t)
\Psi_q^* ({\bf x}', t')
\end{equation}    
with $\Psi_q$ being the Fock states of $H_0$. Thus the nonlinear 
perturbation $H_P$ scatters $\Psi_{a \alpha q}$ into $\Psi_{b \beta p}$,
and vice versa. 
In particular, direct scatterings with
short wavelength modes (environment or noise) would lead to decoherence
of long wavelength modes.
Therefore, long wavelength modes achieve not only classical 
correlation but also decoherence \cite{lombardo} and long wavelength
modes emerge as a classical order parameter.

\section{Conclusion}

The QCD with two massless quarks would undergo a second order phase transition.
The effective theory for the QCD phase transition is
the linear sigma model for the quark-antiquark condensate, that is, 
the sigma and pion fields. 
Under a rapid cooling process, the QCD phase transition proceeds far 
from equilibrium (nonequilibrium). In this talk we 
focused on the dynamical process of the nonequilibrium phase transition 
and its implications on DCC domain growth.

The most prominent feature of nonequilibrium second order 
phase transitions is the instability of long wavelength modes. 
These modes begin to grow exponentially at the onset 
of phase transition while rolling over 
the barrier from the false vacuum to the true vacuum. 
Therefore, this nonequilibrium dynamical process leads to
large domains of disoriented chiral condensate, 
regions of misaligned vacuum. On the contrary, domains formed from
thermal equilibrium have small sizes 
determined by thermal energy and are randomly oriented in isospin space.

Using the nonequilibrium linear sigma model with a smooth finite, 
we observed that DCC domains grow according to some power-law scaling 
relations. We found the power $t^{1/4}$ for the scaling behavior 
in the intermediate stage and the Cahn-Allen scaling power $t^{1/2}$ at 
the later stage of phase transition. 
However, in the Hartree approximation, exponentially growing
long wavelength modes contribute $\lambda \langle \vec{\pi}^2 \rangle$,
which in turn attenuates the instability. The instability completely
stops when  $\lambda \langle \vec{\pi}^2 \rangle$ dominates over $m_f^2$.
Therefore, domains grow for a limited time 
and reach a typical size of $ 1 \sim 2$ fm 
in the Hartree approximation \cite{boyanovsky}. 
This size of DCC domain may not be large enough to 
lead to any significant observation.

It was shown that higher order quantum corrections increase the domains 
size by additional factors $(2 n+1)^{1/2}$ if the duration of 
instability is long enough to make higher order terms grow.  
The longer is the quench time, the larger domains are.  However,
relaxation time scale gives a limitation on domain growth 
through instability, because thermal
equilibration competes with instability
when the quench time is comparable to relaxation time.  
This non-Gaussian effects on DCC domains may lead 
to observations in heavy ion collisions 
and high energy cosmic rays. This mechanism should be 
distinguished from the anomaly enhanced domain formation \cite{asakawa}.

\section*{Acknowledgments}
The author would like to thank K. Rajagopal for useful discussions.
This work is supported by Korea Research Foundation
under grant No. KRF-2003-041-C20053.

\end{document}